\documentstyle[12pt]{article}
\title{Non-equal-time Poisson brackets}
\author{Hrvoje Nikoli\'c  \\
Theoretical Physics Division, Rudjer Bo\v{s}kovi\'{c} Institute, \\
P.O.B. 1016, HR-10001 Zagreb, Croatia \\
{\normalsize hrvoje@faust.irb.hr} \\
\makebox[1in]{} \\
}
\date{\today}
\begin{document}
\maketitle
\begin{abstract}
 The standard definition of the Poisson brackets is generalized to the  
non-equal-time Poisson brackets. 
Their relationship to the equal-time Poisson brackets, as well as to
the equal- and non-equal-time commutators, is discussed.                                     
\end{abstract}

\mbox{ }\newline
There is a well-known correspondence between the Poisson brackets in 
classical mechanics and the equal-time commutators in quantum mechanics.
However, although the generalization to the non-equal-time commutators is    
discussed in every textbook on quantum field theory \cite{bd}, 
we do not know any    
reference where its classical analog -- non-equal-time Poisson brackets --
are explicitly constructed. It seems that this is part of tacit knowledge
among specialists \cite{mar}. In this article we give an explicit 
construction of the non-equal-time Poisson brackets and discuss their
relationship to the equal-time Poisson brackets, as well as to the equal- and   
non-equal-time commutators. 

Let us first discuss a system with one degree of freedom, described by
a Hamiltonian $H(x,p)$. The ordinary Poisson brackets are defined by
\begin{equation}\label{A1}
 \{ A(x,p),B(x,p) \} = \frac{\partial A}{\partial x}\frac{\partial
 B}{\partial p} - \frac{\partial A}{\partial p}\frac{\partial B}{\partial x}
 \; . 
\end{equation}
In particular, 
\begin{equation}\label{A2}
 \{ x,p \}=1 \; , \;\;\;\; \{ x,x \}=\{ p,p \}=0 \; .
\end{equation}
Note that $x$ and $p$ in (\ref{A1}) and (\ref{A2}) have no explicit 
time-dependence. The Poisson brackets do not depend on a particular dynamics,   
i.e. a Hamiltonian. We shall see that these are actually equal-time Poisson
brackets. The variables $x$ and $p$ are classical analogs of quantum operators
in the Schr\"{o}dinger picture.  

Let us discuss the time dependence of the canonical variables. They are
generally functions of the time $t$ and the initial conditions $x(\tau)$,    
$p(\tau)$, where $\tau$ is some {\em fixed} instant. Thus we introduce 
the notation
\begin{equation}\label{A3}
 x_{G}(t)=x(t,x(\tau),p(\tau)) \: ,
\end{equation}
and similarly for $p_{G}(t)$,
where the label $G$ indicates that this is the general solution of the equations
of motion. If the equations of motion are linear, then $x_{G}(t)$ and  
$p_{G}(t)$ are linear in $x(\tau)$ and $p(\tau)$.
We define the derivative
\begin{equation}\label{A4}
 \frac{\partial F(x(t_i))}{\partial x(t_j)} \equiv
 \lim_{\varepsilon \rightarrow 0^{+} } 
  \int_{t_j -\varepsilon}^{t_j +\varepsilon}
  ds \frac{\delta F(x(t_i))}{\delta x(s)} =
  \lim_{\varepsilon \rightarrow 0^{+} } \int_{t_j -\varepsilon}
  ^{t_j +\varepsilon} ds\: F'(x(t_i)) \: 
  \delta (t_i -s) = \delta_{t_i , t_j} \: F'(x(t_i))
\end{equation}
and the non-equal-time Poisson bracket
\begin{equation}\label{A5}
 \{ A(x_G(t),p_G(t)),B(x_G(t'),p_G(t')) \}_{\tau} = 
 \frac{\partial A}{\partial x(\tau)}\frac{\partial
 B}{\partial p(\tau)} - \frac{\partial A}{\partial p(\tau)}
 \frac{\partial B}{\partial x(\tau)} \; .
\end{equation}
The right-hand side of (\ref{A5}) can also be calculated if
$x_G(t)$ is replaced by $x(t)$ or $p_G(t)$ by $p(t)$. For example, 
\begin{eqnarray}\label{A6}   
 & \{ x(t),p(t') \}_{\tau}=\delta_{t,t'}\delta_{t,\tau} \; , & \nonumber \\ 
 & \{ x(t),x(t') \}_{\tau}=\{ p(t),p(t') \}_{\tau}=0 \; . &
\end{eqnarray}
The ordinary Poisson bracket (\ref{A1}) can be obtained from (\ref{A5})
using the relation 
\begin{equation}\label{A7}
 \{ A(x_G(\tau),p_G(\tau)),B(x_G(\tau),p_G(\tau)) \}_\tau =
 \{ A(x,p),B(x,p) \} \; 
\end{equation} 
and the identifications $x(\tau)\equiv x$ and $p(\tau)\equiv p$.   
The transition from the non-equal-time Poisson brackets in classical
mechanics to the non-equal-time commutators in the Heisenberg
representation of quantum physics is given by
\begin{equation}\label{A8}
 \{ A(x_G(t),p_G(t)),B(x_G(t'),p_G(t')) \}_{\tau =0} \longrightarrow -i
 [ A(\hat{x}(t),\hat{p}(t)),B(\hat{x}(t'),\hat{p}(t')) ] \; .
\end{equation}
If $A$ and $B$ are linear functions and if the equations of motion are linear,
then relation (\ref{A8}) is an equality. In a more general case, this does not
need to be an equality because of the ordering ambiguities of quantum
operators. 

Let us illustrate all this on a simple example. We consider the Hamiltonian  
\begin{equation}\label{A9} 
 H(x,p)=\frac{p^2}{2} + \frac{x^2}{2} \; .
\end{equation}
The general solution of the corresponding equations of motion is
\begin{equation}\label{A10}
 x_G(t)=a e^{-it} + a^* e^{it} \; ,
\end{equation}
which we write in the form
\begin{equation}\label{A11}
 x_G(t)= x(\tau) \cos (t-\tau) + p(\tau) \sin (t-\tau) \; .
\end{equation}
(There is a simple relationship between the $(a,a^*)$ and $(x(\tau),p(\tau))$
coefficients.)
The corresponding canonical momentum is 
\begin{equation}\label{A12}
 p_G(t)=\dot{x}_G(t)=
 -x(\tau) \sin (t-\tau) + p(\tau) \cos (t-\tau) \; . 
\end{equation}
From (\ref{A5}) we obtain 
\begin{equation}\label{A13}
 \{x_G(t),p_G(t')\}_{\tau} =
 \cos(t-\tau)\cos(t'-\tau) + \sin(t-\tau)\sin(t'-\tau) 
\end{equation}
and 
\begin{equation}\label{A14}
  \{x_G(t),p_G(t)\}_{\tau} = 1 \; ,
\end{equation}
as a special case of (\ref{A13}). On the other hand, in quantum mechanics
we work with the operators in the Heisenberg picture $\hat{x}(t)$,
$\hat{p}(t)$ and the  
corresponding operators in the Schr\"{o}dinger picture $\hat{x}=\hat{x}(0)$,
$\hat{p}=\hat{p}(0)$, which satisfy
\begin{eqnarray}\label{A15}
 & \hat{x}(t)=\hat{x}\cos t + \hat{p}\sin t \; , & \nonumber \\
 & \hat{p}(t)=-\hat{x}\sin t + \hat{p}\cos t \; . &
\end{eqnarray}
From the equal-time commutation relations
\begin{equation}\label{A16}
 [\hat{x},\hat{p}]=i \; , \;\;\;\; [\hat{x},\hat{x}]=[\hat{p},\hat{p}]=0
\end{equation}
we find the non-equal-time commutation relation
\begin{equation}\label{A17}
 [ \hat{x}(t),\hat{p}(t')]=i(\cos t \:\cos t' + \sin t \:\sin t') \; .
\end{equation}
The equality (\ref{A8}) is obtained by putting $\tau =0$ in (\ref{A13}).

The generalization of the non-equal-time Poisson brackets
to a discrete set of degrees of freedom
is trivial. Let us shortly discuss the generalization to field theory.
The ordinary (i.e., equal-time) Poisson bracket is
\begin{equation}\label{A1f}
 \{ A(\phi({\bf x}),\pi({\bf x})),B(\phi({\bf x'}),\pi({\bf x'})) \} =   
 \int d^{3} y \left[ \frac{\delta A}{\delta\phi({\bf y})}
 \frac{\delta B}{\delta\pi({\bf y})} - 
 \frac{\delta A}{\delta\pi({\bf y})} \frac{\delta B}{\delta\phi({\bf y})}
 \right] \; . 
\end{equation}
In particular, 
\begin{eqnarray}\label{A2f}
 & \{ \phi({\bf x}),\pi({\bf x'}) \}=\delta^{3}({\bf x}-{\bf x'})
  \; , & \nonumber \\ 
 & \{ \phi({\bf x}),\phi({\bf x'}) \}=\{ \pi({\bf x}),\pi({\bf x'}) \}=0 \;
.
\end{eqnarray}
Now we introduce the space-time point $x=(x_0,{\bf x})$.
The generalization of (\ref{A4}) is
\begin{eqnarray}\label{A4f}
  \frac{\partial F(\phi(x))}{\partial \phi(y)} & =
 & \frac{\partial F(\phi(x_0,{\bf x}))}{\partial \phi(y_0,{\bf y})}  \equiv
 \lim_{\varepsilon \rightarrow 0^{+} } \int_{y_0 -\varepsilon}
  ^{y_0 +\varepsilon} ds \:\frac{\delta F(\phi(x_0,{\bf x}))}
 {\delta \phi(s,{\bf y})}  \nonumber \\ 
  & = & \lim_{\varepsilon \rightarrow 0^{+} } \int_{y_0 -\varepsilon}
  ^{y_0 +\varepsilon} ds \: F'(\phi(x_0,{\bf x})) 
 \:\delta^{3}({\bf x}-{\bf y}) \delta(x_0 -s) \nonumber \\ 
 & = & 
 \delta^{3}({\bf x}-{\bf y})\delta_{x_0,y_0}\: F'(\phi(x_0,{\bf x}))   
\end{eqnarray}
and the generalization of (\ref{A5}) is
\begin{equation}\label{A5f}
 \{ A(\phi_G(x),\pi_G(x)),B(\phi_G(x'),\pi_G(x')) \}_{\tau} =
 \int d^{3}y \left[ \frac{\partial A}{\partial\phi(\tau,{\bf y})}
 \frac{\partial B}{\partial\pi(\tau,{\bf y})} -
 \frac{\partial A}{\partial\pi(\tau,{\bf y})}
 \frac{\partial B}{\partial\phi(\tau,{\bf y})} \right] \; .
\end{equation}
As in the case of one degree of freedom, 
we have the correspondence between classical and quantum field 
theory in the form
\begin{equation}\label{A8f}
 \{ A(\phi_G(x),\pi_G(x)),B(\phi_G(x'),\pi_G(x')) \}_{\tau =0}
 \longrightarrow -i
 [ A(\hat{\phi}(x),\hat{\pi}(x)),B(\hat{\phi}(x'),\hat{\pi}(x')) ] \; ,
\end{equation}
which is an equality for a linear case. 

In this article we have constructed the non-equal-time Poisson brackets by
a generalization of the standard definition of the Poisson brackets. It is
interesting
to note that in \cite{pei} the non-equal-time Poisson brackets for field 
theory are constructed in a completely different way, directly from
Lagrangians in a manifestly covariant way. However, this construction
does not coincide with the conventional construction (such as ours), because
the antisymmetry of the Poisson brackets is not generally provided in the 
approach of \cite{pei}. However, this construction still does coincide with the 
conventional approach for a large class of Lagrangians. 

\section*{Acknowledgement}

This work was supported by the Ministry of Science and Technology of the
Republic of Croatia under Contract No. 00980102.

\end{document}